# Cosolvency response in polymer brushes

Huaisong Yong[1, 2, 3]\*, Binyu Zhao[4, 5]

[1]School of New Energy and Materials, Southwest Petroleum University, Chengdu 610500, China

[2]Department of Molecules & Materials, MESA+ Institute, University of Twente, Enschede 7500 AE, The Netherlands

[3]Institute Theory of Polymers, Leibniz-Institut für Polymerforschung Dresden e.V., Dresden D-01069, Germany

[4]Key Laboratory of Green and High-End Utilization of Salt Lake Resources, Qinghai Institute of Salt Lakes, Chinese Academy of Sciences, Xining 810008, China

[5]Qinghai Provincial Key Laboratory of Resources and Chemistry of Salt Lakes, Xining 810008, China

\*Correspondence: yonghs@swpu.edu.cn, yonghuaisong@gmail.com

## 0. Abstract

**Hypothesis and Theory**

We present the first analytic theory with elegant and closed-form analytical solutions to explore the cosolvency effect in polymer brushes, where polymer chains that are poorly soluble in two pure solvents become fully soluble in certain mixtures thereof. This effect is key to designing stimulus-responsive smart materials but has not previously been addressed by analytic theory for polymer brushes.

**Findings**

Our theoretical framework reveals that preferential adsorption of cosolvent induces an effective repulsion between monomers solvated by cosolvent and those solvated by solvent. The equilibrium solvation of polymer chains by cosolvent gives rise to a concentration-dependent $\chi$-function, which captures the effective interactions within the brush and reproduces the reentrant behavior characteristic of the cosolvency effect. The model predicts a discontinuous soluble transition followed by a re-collapse transition at higher cosolvent concentrations. Analytical treatment within a minimal free-energy model for the case of two symmetric poor solvents shows that the swelling and re-collapse transitions share the same thermodynamic origin. For low-density brushes, we derive an analytical approximation and delineate the phase diagram of parameter space in which discontinuous transitions occur. For cosolvency to take place, the theory specifies a minimum strength for preferential solvation and the associated repulsive coupling. Furthermore, it demonstrates that, contrary to previous models, repulsive interactions between cosolvent and solvent in the bulk are not required. This work lays the groundwork for the rational design of smart stimulus-responsive materials based on the cosolvency effect in polymer brushes, a capability which was not previously established.

**Graphical abstract**

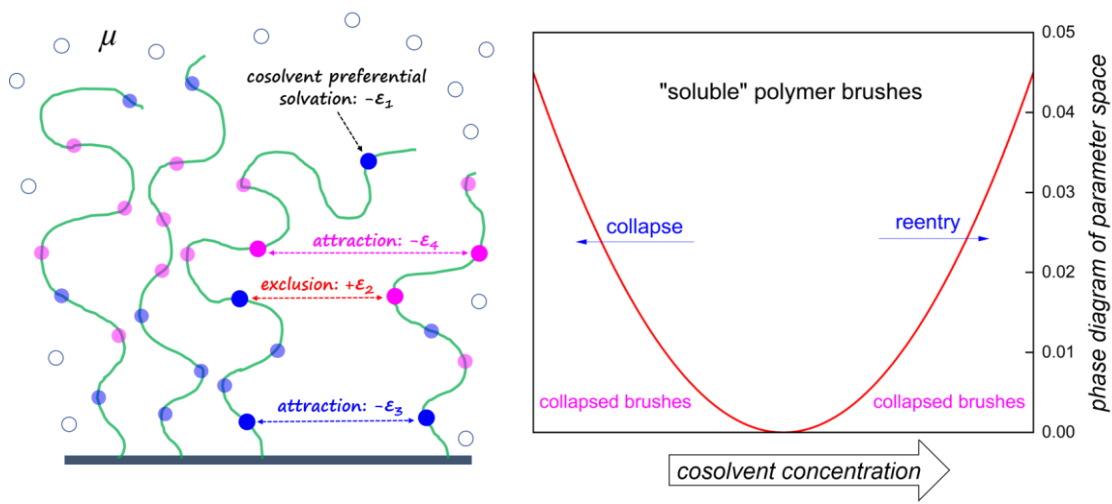

## 1. Introduction

A wealth of phase behaviors can be exhibited when polymers are mixed with multi-component solvents [1-7], such as reentrant condensation [1, 5, 6]. A counterintuitive example of reentrant condensation is that a polymer poorly soluble in two pure solvents can become completely soluble in certain mixtures of these two poor solvents, a phenomenon well known as the cosolvency effect [5]. Cosolvency effect was reported experimentally for polymer materials as early as over half a century ago [8]; however, it remains enigmatic and poorly understood. We note that cosolvency has been reported almost exclusively for polymers in aqueous solution, such as in mixtures of alcohols and water [9-13]. This, however, does not mean that cosolvency cannot occur in non-aqueous solutions [14-18]. For example, poly(methyl methacrylate) [16] exhibits cosolvency in binary mixtures of two organic solvents (1-propanol and cyclohexane).

A fact worthy of note is that poly(methyl methacrylate) also shows cosolvency effect in aqueous solution of alcohols [13]. Remarkably, different polymers, such as thermo-responsive polysiloxanes [15], can display cosolvency both in binary mixtures of organic solvents and in aqueous solutions containing an organic solvent. These observations indicate that cosolvency does not necessarily arise from unique properties of a specific polymer or solvent. This suggests that it may be sufficient to understand cosolvency based solely on non-specific interactions among the polymer and the two poor solvents. These facts motivated pioneering works **[19-23]** to extend classical Flory–Huggins (FH) theory **[24, 25]** using non-specific interactions, and it was found that a repulsive interaction between the two poor solvents in the bulk can lead to cosolvency.

Nevertheless, these extended FH theories **[19-23]** assumed purely isotropic interactions and homogeneous distribution of all components in polymer solutions, assumptions that cannot fully explain previous experimental and simulation studies on cosolvency. For example, both molecular dynamics simulations **[26-29]** and experiments **[9-11, 14, 17, 18, 28, 30-32]** indicated that cosolvency can arise from the relatively preferential solvation of heterogeneous-like polymers by one solvent, as observed in homopolymers **[9, 14]**, alternating **[10]** and random **[11]** copolymers in binary mixtures of two poor solvents. We recognize that such preferential solvation inevitably leads to relative enrichment of one solvent around or on polymer chains, as reported in previous experiments **[17, 18, 28, 30-32]**. We also note that in polymer solutions, most of the solution volume is occupied by solvent molecules rather than polymer chains. These facts imply that the influence of interactions between the two poor solvents on polymer phase transitions may be overestimated by previous extended FH theories **[19-23]**, especially when solvent molecules are far from polymer chains (i.e., in the bulk). Therefore, in formulating a theory, it is necessary to incorporate the effect of preferential solvation while mitigating the overestimation of solvent–solvent interactions, an approach which remains absent in the current literature. Moreover, in contrast to polymer solutions, only experimental studies **[17, 33-36]** but no theory have investigated cosolvency in polymer brushes, where polymers adopt brush-like conformations and consist of chains end-anchored to a surface, as illustrated schematically in **Figure 1**.

Thus, the primary goal of this work is to formulate the first analytic theory with closed-form analytical solutions to explore the cosolvency effect in polymer brushes. We systematically analyze the role of preferential solvation in determining the phase behavior of cosolvency. Based on this analysis, we propose a mean-field model that captures the essential physics and reproduces experimentally observed trends. In the remainder of this article, a mean-field theory for polymer brushes in terms of free energy will be constructed in **Section 2**. From **Section 3** to **Section 6**, its analytical solution will be examined in detail, and simplified phase diagrams for polymer brushes will be presented. In these sections, we outline general results of the theory and, where appropriate, discuss their applicability. We address the novelty and limitations of our theory in **Section 7**, and conclude with a summary of main findings in **Section 8**.

## 2. Free energy model for cosolvency response in polymer brushes

In general, the sizes of a monomer and a solvent molecule can differ significantly. Since we are interested in the general physical understanding of the model, we restrict ourselves here to the symmetric case, i.e., we consider these sizes to be identical in the spirit of the classical Flory–Huggins model **[24, 25]**. Here and in the following, we consider the energy per monomer in units of $k_B T$ for an incompressible system, unless otherwise specified.

We note that previous experimental **[9-11, 14, 17, 18, 28, 30-32]** and simulation **[26-29]** studies have indicated that polymer chains can be relatively and preferentially solvated by one of the two poor solvents during cosolvency. Here, preferential solvation represents a form of anisotropic interaction, which may originate from interactions such as hydrogen bonds **[32]** or dipole interactions **[37, 38].** As illustrated in **Figure 1**, this observation prompts us to account for the preferential solvation effect on or near polymer chains through a *one-dimensional* mixing free energy, $g_{ads}$,

$$g_{ads} = \varphi \ln(\varphi) + (1-\varphi)\ln(1-\varphi) - \mu\varphi - \varepsilon_1\varphi + \chi_s\varphi(1-\varphi) \tag{1}$$

Here, we term the preferentially adsorbed poor solvent as the cosolvent, and refer to the other poor solvent simply as the solvent for convenience. The fraction of monomers preferentially solvated by the cosolvent is denoted by $\varphi$. The preferential-solvation strength per cosolvent molecule with respect to the polymer is denoted by $\varepsilon_1$ in units of $k_B T$. The exchange chemical potential associated with transferring a cosolvent molecule from the bulk solvent mixture to the polymer chains is denoted by $\mu$. In this study, it is not necessary to consider a specific form of $\mu$ unless otherwise specified.

The demixing tendency between the two poor solvents may also influence the cosolvency effect of a polymer. For instance, our previous experiments **[11]** showed that the demixing tendency between alcohol and water has a noticeable effect on the cosolvency behavior of certain copolymers. We account for the effect of demixing tendency near or on the brush chains via the parameter $\chi_s$, which describes the non-ideal one-dimensional mixing of cosolvent and solvent on the polymer chains. This parameter includes contributions from both excess enthalpy and excess entropy. In this work, we assume that no phase separation occurs between the two poor solvents during cosolvency, which requires $\chi_s < 2$. This condition corresponds to the only scenario reported in experiments to date.

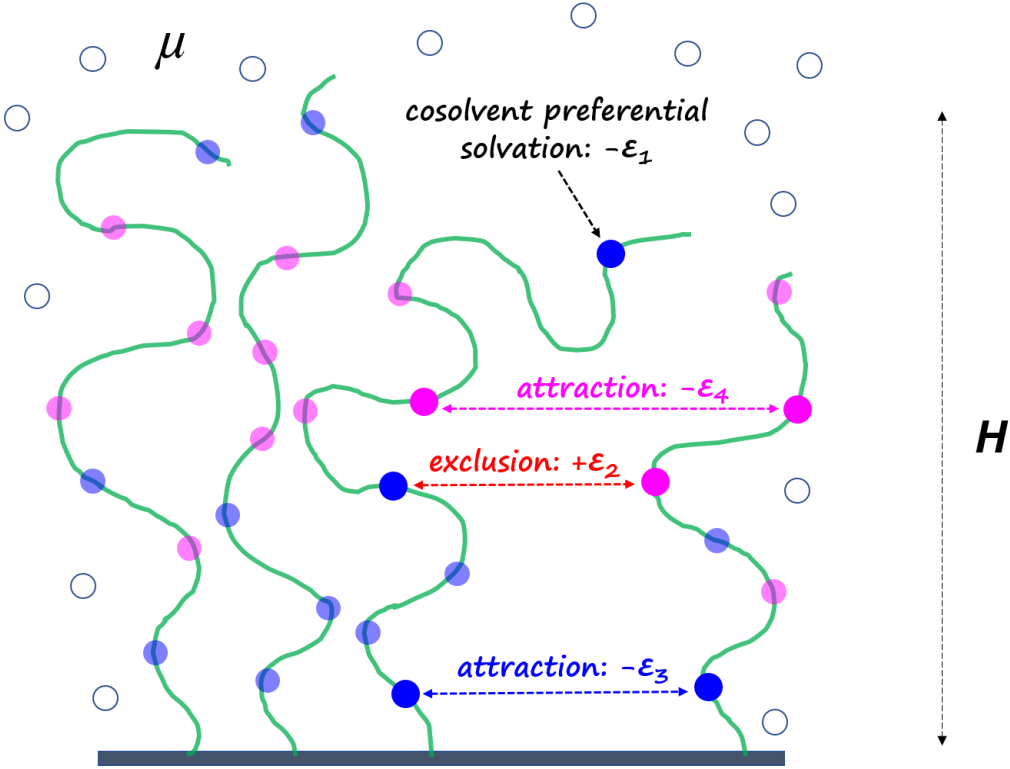

**Figure 1.** A schematic illustration of the preferential-solvation effect of cosolvent on polymer brush chains. In the figure, green lines represent polymer chains, filled blue circles represent adsorbed cosolvent, and filled pink circles represent adsorbed common solvent. Dashed arrows indicate non-specific short-range interactions. For simplicity, non-adsorbed solvent molecules in the background are represented by open black circles.

To capture the complete solubilization of a polymer in a mixture of two poor solvents in cosolvency effect, we propose that it is necessary to consider an effective repulsion between cosolvent-solvated and solvent-solvated monomers. This repulsion, arising from the preferential solvation of polymer chains by the cosolvent, can be statistically described by $g_{rep}$,

$$g_{rep} = 2\varepsilon_2 \varphi(1-\varphi)c \qquad (2)$$

We denote the repulsion strength per adsorbed cosolvent molecule as $\varepsilon_2$, and denote the volume fraction of the polymer as $c$. For heterogeneous-like polymers **[9-11, 14, 17, 18, 28, 30-32]**, the repulsion effect can also be understood by considering that if two different solvent molecules adsorb onto the polymer chains, the monomers solvated by these two solvents will effectively repel each other.

The attraction effect between cosolvent-solvated monomers, as well as between solvent-solvated monomers, is described by $g_{attr}$, which shall satisfy the limiting condition that the solvent quality is poor for both pure solvents if $\varphi \to 0$ and $\varphi \to 1$.

$$g_{attr} = -\varepsilon_3 \varphi^2 c - \varepsilon_4 (1-\varphi)^2 c \tag{3}$$

Here, we denote the effective attraction between polymers when immersed in the cosolvent as $\varepsilon_3$, and the effective attraction between polymers when immersed in the pure solvent as $\varepsilon_4$. To recover the poor solvent condition in the limits $\varphi \to 0$ and $\varphi \to 1$, the boundary conditions $\varepsilon_3 > 1/2$ and $\varepsilon_4 > 1/2$ must hold.

The entropic contribution to the free energy arising from the *three-dimensional* mixing of brushes with the two poor solvents is given by $g_{brush}$,

$$g_{brush} = \frac{1}{2}\left(\frac{\sigma}{c}\right)^2 + \left(\frac{1}{c}-1\right)\ln(1-c) \tag{4}$$

The first term of $g_{brush}$ represents the elastic entropic free energy of the brushes, which results from chain elasticity and the absence of translational entropy of the brush chains. We describe the chain elasticity using the well-established classical Alexander–de Gennes approach **[39, 40]** for uncharged brushes **[41]**, whose simplicity also facilitates our subsequent analytical calculations. Here, the grafting density is denoted by $\sigma$, subject to the physical constraint $0 < \sigma \leq 1$. We assume that the two poor solvents mix ideally when far from the polymer chains, which leads to the second term of $g_{brush}$.

The overall free energy per monomer in the framework of the *NVT* ensemble is given by $g(\varphi, c) = g_{ads} + g_{rep} + g_{attr} + g_{brush}$. One essential feature of our free-energy construction is that enthalpic effects dominate the phase transition, this will be clearly seen from the solution of our model from **Section 3** to **Section 6**. We note that this is coincident with recent simulation observations by Zhou and coworkers **[26]** to explain the cosolvency of cyanoethyl cellulose in mixtures of acetone and water. Another essential feature of our free-energy construction is that we focus on the solvation, repulsion, and attraction effects arising from anisotropic interactions between the two solvents near or on the polymer chains, while neglecting non-essential mixing effects when solvent molecules are far from the chains. This approach helps mitigate the overestimation of the role of solvent–solvent interactions in the bulk, without sacrificing the generality needed to obtain key physical insights. We note that our theoretical formulation for cosolvency differs qualitatively from earlier models proposed by Meng **[19, 21]**, Zhang **[20]**, Xiao **[22]**, and Dudowicz **[23]** with their respective coworkers. Those works assumed purely isotropic interactions in the spirit of the classical Flory–Huggins model **[24, 25]** and showed that a repulsive effect between the two poor solvents in the bulk, rather than near or on the polymer chains, can lead to the cosolvency effect.

## 3. Maximum coupling approximation and non-monotonic effective Flory–Huggins $\chi$ parameter

According to the scaling theory for polymer brushes **[39, 40, 42]**, a polymer brush becomes more difficult for a phase transition to occur as its grafting density increases. This is evident in the limiting case of grafting density $\sigma = 1$, where the brush phase is fully occupied by completely stretched chains and impossibly making a sudden change in brush thickness. Therefore, for a phase transition to occur in polymer brushes, the parameters $\varepsilon_2$, $\varepsilon_3$, and $\varepsilon_4$ must satisfy a grafting density-dependent relation. In this subsection, we derive this relation at the mean-field level. To simplify the analytical treatment in this and subsequent sections, we primarily consider the symmetric case $\varepsilon_3 = \varepsilon_4 > 1/2$ unless otherwise noted. This simplification does not alter the general physical conclusions of the model.

To fully understand the phase behavior of polymer brushes as a function of the cosolvent volume fraction, one must minimize the free energy $g(\varphi, c)$ with respect to $\varphi$. Substituting the solution $\varphi(\mu, c)$ back into the free energy yields an effective free energy for the polymer, in which the preferential-solvation effect of the cosolvent is mapped onto an effective monomer–monomer interaction that depends on the cosolvent concentration. We find that when only the preferential solvation between cosolvent and polymer is considered, linear polymers effectively behave as one-dimensional substrates for cosolvent and solvent molecules. Theoretically, this situation can be treated in formal analogy with the one-dimensional Ising model **[43,**

**44]**. Hence, one should not expect a phase transition to arise solely from the simple exchange of cosolvent and solvent adsorption on polymer chains. This indicates that, in our model, rather than by other factors, the phase transition of polymer brushes in solvent mixtures is governed by the repulsive coupling between cosolvent-solvated and solvent-solvated monomers, i.e., the term $g_{rep}$ in our free-energy model. In other words, we can exploit this feature to simplify the analytical solution of our model.

Based on the construction of $g_{rep}$ and $g_{attr}$, we see that the maximum coupling is achieved at $\varphi = 1/2$, where the maximum expansion of polymer chains is expected for the symmetric case $\varepsilon_3 = \varepsilon_4$. This feature significantly simplifies our analytical calculations by introducing a perturbation $\delta$ from the half-occupation of the chain by the cosolvent according to

$$\varphi = \frac{1}{2}(1-\delta) \tag{5}$$

This perturbation approach is analogous to that used in our previous works **[45-48]** and was numerically validated by Sommer **[49]** recently. Here, $\delta > 0$ corresponds to the phase transition from insoluble to soluble states (soluble transition), and $\delta < 0$ corresponds to the phase transition from soluble to insoluble states (reentry transition). By performing a Taylor expansion of the $\delta$-dependent terms in the logarithm function up to cubic terms ($\delta^3$), under the constraint that quartic terms satisfy $\delta^4 \ll 1$, we obtain

$$g(c,\delta) = \frac{(\mu+\varepsilon_1)}{2}\delta + \frac{1}{2}\delta^2 + \frac{1}{2}\varepsilon_2(1-\delta^2)c - \frac{1}{2}\varepsilon_3(1+\delta^2)c \\ + \frac{1}{4}\chi_s(1-\delta^2) - \frac{1}{2}(\mu+\varepsilon_1+\ln 4) + g_{brush} \tag{6}$$

Note that the cubic terms ($\delta^3$) cancel out in the Taylor expansion and do not appear in **Equation (6)**.

The equilibrium condition with respect to $\delta$ by $\partial g/\partial \delta = 0$ leads to

$$\delta = -\frac{1}{2}\left[\frac{(\mu+\varepsilon_1)}{1-\frac{1}{2}\chi_s - (\varepsilon_2+\varepsilon_3)c}\right] \tag{7}$$

Resubstituting this result into **Equation (6)**, we obtain the free energy of the polymer brushes at a given grafting density at equilibrium with the cosolvent concentration at

$$g(c) = -\frac{1}{2}\left(\mu + \varepsilon_1 + \ln 4 - \frac{1}{2}\chi_s\right) + \frac{(\varepsilon_2 - \varepsilon_3)}{2}c - \frac{1}{8}\frac{(\mu + \varepsilon_1)^2}{\left[1 - \frac{1}{2}\chi_s - (\varepsilon_2 + \varepsilon_3)c\right]} + g_{brush}$$

$$= -\frac{1}{2}\left(\mu + \varepsilon_1 + \ln 4 - \frac{1}{2}\chi_s\right) - \frac{1}{8}\frac{(\mu + \varepsilon_1)^2}{\left(1 - \frac{1}{2}\chi_s\right)} - \chi_{eff}c + g_{brush} \qquad (8)$$

with the effective Flory parameter $\chi$ function

$$\chi_{eff} = \frac{1}{2}(\varepsilon_3 - \varepsilon_2) + \frac{(\mu + \varepsilon_1)^2}{8} \frac{(\varepsilon_2 + \varepsilon_3)}{\left(1 - \frac{1}{2}\chi_s\right)\left[1 - \frac{1}{2}\chi_s - (\varepsilon_2 + \varepsilon_3)c\right]} \qquad (9)$$

From the expression of **Equation (9)**, we see that a typical choice for parameters $0 < \varepsilon_2 < 1/2$ and $\varepsilon_3 > 1/2$ with $\varepsilon_2 + \varepsilon_3 \leq 1 - \chi_s/2$, yields a minimum value of the $\chi$-function in the range $0 < \chi_{eff,0} = (\varepsilon_3 - \varepsilon_2)/2 < 1/2 - \chi_s/4 \leq 1/2$ at the point $\mu = -\varepsilon_1$, corresponding to the half-solvated state of the polymers by the cosolvent. If the chemical potential $\mu$ is far away from the point $\mu = -\varepsilon_1$, the $\chi$-function will increase significantly with very low or high concentrations of cosolvent (corresponding to the large value of $|\mu|$). We see that the construction of **Equation (9)** recovers the typical reentrant behaviors of cosolvency effect. It is also noted that if the chemical potential $\mu$ is far away from the point $\mu = -\varepsilon_1$, the $\chi$-function will change with variation of polymer concentration which forbids a simple interpretation for a $\chi$-parameter as in the classical Flory-Huggins model [24, 25]. In fact, our model gives rise to a modification of all virial coefficients, in contrast to the classical Flory-Huggins model [24, 25] where only the second virial coefficient plays an important role. We will discuss in detail the virial-expansion of our theory in **Section 5**.

## 4. Construction of analytical phase diagrams

In order to analytically discuss the phase transition, the net osmotic pressure $\Pi(c, \mu; \sigma, \varepsilon_1, \varepsilon_2, \varepsilon_3)$ in polymer brushes can be calculated from **Equation(8)** as

$$\Pi = c^2 \frac{\partial g}{\partial c} = \frac{(\varepsilon_2 - \varepsilon_3)}{2}c^2 - \frac{(\mu + \varepsilon_1)^2}{8}\frac{(\varepsilon_2 + \varepsilon_3)c^2}{\left[1 - \frac{1}{2}\chi_s - (\varepsilon_2 + \varepsilon_3)c\right]^2} - \frac{\sigma^2}{c} - \ln(1 - c) - c \qquad (10)$$

Both types of solvent molecules can diffuse across the boundary of the polymer brushes. This means that when the polymer brushes are in equilibrium, the balance of mechanical forces within the brushes requires that the net osmotic pressure must be zero, $\Pi = 0$ (e.g., see a recent excellent monograph by Muthukumar **[50]** on this topic). This fundamental law leads to

$$\frac{(\mu+\varepsilon_1)^2}{8} = U^2 = \frac{\left[1-\frac{1}{2}\chi_s-(\varepsilon_2+\varepsilon_3)c\right]^2}{(\varepsilon_2+\varepsilon_3)}\left[\frac{\varepsilon_2-\varepsilon_3}{2}-\frac{\sigma^2}{c^3}-\frac{\ln(1-c)}{c^2}-\frac{1}{c}\right] \tag{11}$$

Here, we denote the external adsorption field $U$ as $U = (\mu + \varepsilon_1)/2\sqrt{2}$. We recognize that the term $U^2$ represents the square of a function of the cosolvent density, yielding two solutions for the chemical potential $\mu$. This symmetry indicates that a reentrant transition can occur for polymer brushes in the presence of a cosolvent, with soluble and insoluble transitions at low and high cosolvent concentrations, respectively.

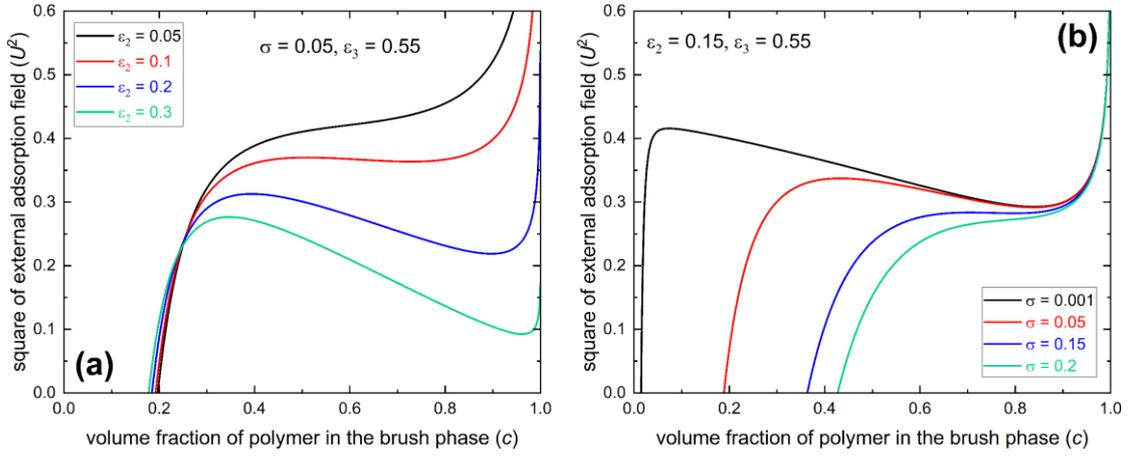

**Figure 2.** Based on **Equation(11)**, the square of the external adsorption field, $U^2$, is plotted as a function of the monomer volume fraction ($c$) for the model parameters $\varepsilon_2 + \varepsilon_3 \leq 1 - \chi_s/2$, with $\varepsilon_3 = 0.55$ and $\chi_s = 0$. In panel (a): illustrating the effect of repulsion between cosolvent-solvated and solvent-solvated monomers ($\varepsilon_2$) on the phase transition in polymer brushes for $\sigma = 0.05$. In panel (b): illustrating the effect of grafting density ($\sigma$) on the phase transition for $\varepsilon_2 = 0.15$.

For the general case of a discontinuous phase transition, the square of the external adsorption field ($U^2$) must exhibit an unstable region of negative compressibility, given by $\partial U^2/\partial c < 0$. In **Figure 2** and **Figure 3**, we respectively display "$U^2 - c$" curves for the conditions $\varepsilon_2 + \varepsilon_3 \leq 1 - \chi_s/2$ and $\varepsilon_2 + \varepsilon_3 > 1 - \chi_s/2$. The coexistence region of the binodal can be obtained via the Maxwell construction using the equal-area criterion. This cannot be derived analytically in an exact manner and must be computed numerically; an example is shown in **Figure 4a**. Nevertheless, we can draw some qualitative conclusions from **Figure 2a** and **Figure 3a**. It is clear that the monomer density in the dilute phase decreases with increasing repulsion strength ($\varepsilon_2$); however, the monomer density in the condensed phase follows a different trend. We also observe that the monomer density in the condensed phase increases with larger $\varepsilon_2$ under the condition $\varepsilon_2 + \varepsilon_3 \leq 1 - \chi_s/2$, while it decreases with larger $\varepsilon_2$ when $\varepsilon_2 + \varepsilon_3 > 1 - \chi_s/2$. We note that these contrasts are somehow consistent with experimental results reported by Cowie and McEwen **[51]** for the cosolvency effect of polymer solutions. Nevertheless, the exact mechanism behind these contrasts is currently unknown and will be discussed in detail in an upcoming study of ours.

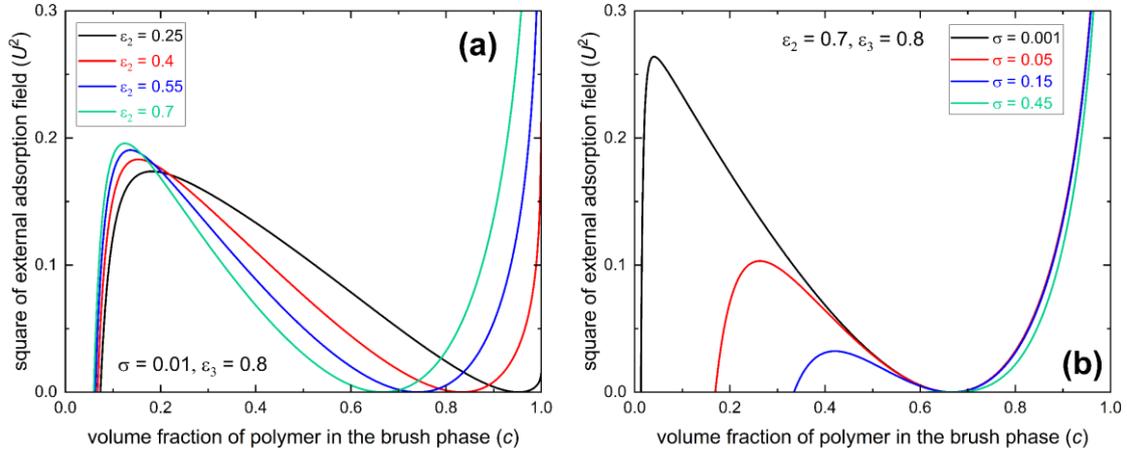

**Figure 3.** Based on **Equation(11)**, the square of the external adsorption field, $U^2$, is plotted as a function of the monomer volume fraction ($c$) for the model parameters $\varepsilon_2 + \varepsilon_3 > 1 - \chi_s/2$, with $\varepsilon_3 = 0.80$ and $\chi_s = 0$. In panel (a): illustrating the effect of repulsion between cosolvent-solvated and solvent-solvated monomers ($\varepsilon_2$) on the phase transition in polymer brushes for $\sigma = 0.01$. In panel (b): illustrating the effect of grafting density ($\sigma$) on the phase transition for $\varepsilon_2 = 0.70$.

We are able to calculate the spinodal state analytically, at which the brushes begin to become unstable, whose existence is a necessary condition for a discontinuous transition scenario. The spinodal of the polymer brushes is given by $\partial U^2/\partial c = 0$ and can be written in the following form

$$0 = \frac{\partial U^2}{\partial c} = -2\left[1 - \frac{1}{2}\chi_s - (\varepsilon_2 + \varepsilon_3)c\right]\left[\frac{\varepsilon_2 - \varepsilon_3}{2} - \frac{\sigma^2}{c^3} - \frac{\ln(1-c)}{c^2} - \frac{1}{c}\right]$$
$$+ \frac{\left[1 - \frac{1}{2}\chi_s - (\varepsilon_2 + \varepsilon_3)c\right]^2}{(\varepsilon_2 + \varepsilon_3)}\left[\frac{3\sigma^2}{c^4} + \frac{2-c}{(1-c)c^2} + \frac{2\ln(1-c)}{c^3}\right]$$

(12)

In general, a "$U^2 - c$" curve can have two physically meaningful spinodal points when the grafting densities of the polymer brushes are moderate or low. This can be clearly observed in **Figure 2b** and **Figure 3b**. From **Equation(12)**, we also recognize that under the condition $\varepsilon_2 + \varepsilon_3 > 1 - \chi_s/2$, the $U^2$ curve is tangent to the horizontal axis, and the abscissa of the tangency point in **Figure 3** is given by $c_{plain} = (1 - \chi_s/2)/(\varepsilon_2 + \varepsilon_3)$. We get a plain spinodal point ($\mu = \varepsilon_1$, $c_{plain}$) by inserting $c_{plain}$ into **Equation(11)**, which is however dynamically inaccessible due to strong fluctuation effects of polymer density in brushes **[52-54]**. In **Figure 4b**, we show the effect of the demixing tendency between two poor solvents on polymer chains ($\chi_s$) on the square of the external adsorption field $U^2$. Our general observation from the "$U^2 - c$" curves is that the phase transition becomes easier with a stronger demixing tendency ($\chi_s$). We note that this analytical prediction is consistent with our previous experimental observation **[11]** on the cosolvency of a copolymer in a series of binary mixtures of alcohols and water.

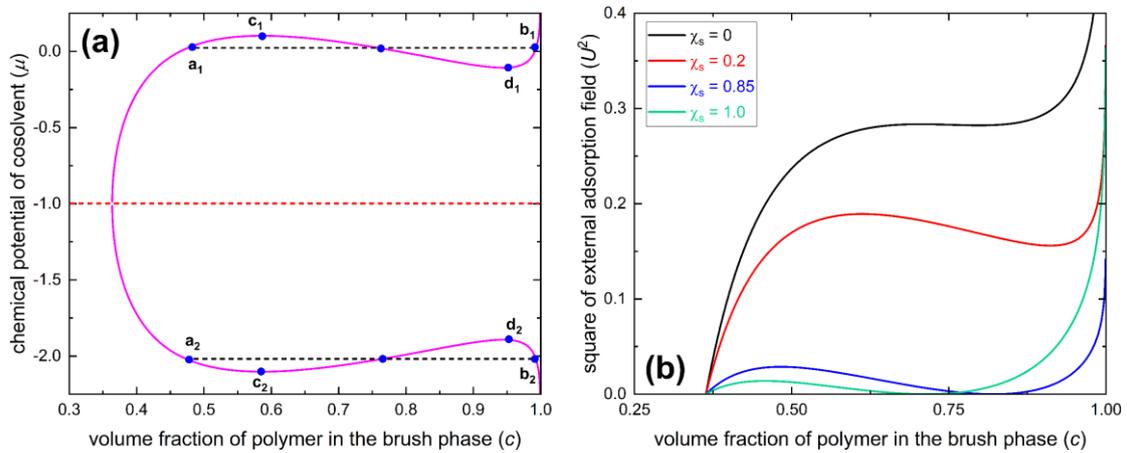

**Figure 4.** Based on **Equation(11)**, the external adsorption field $U$ is plotted as a function of the monomer volume fraction ($c$) for $\varepsilon_2 = 0.15$, $\varepsilon_3 = 0.55$, and $\sigma = 0.15$. In panel (a): an example of using the Maxwell construction to determine the true equilibrium states in the phase transition of polymer brushes for the model parameters $\chi_s = 0.3$ and $\varepsilon_1 = 1$. The equilibrium states in panel (a) are indicated by filled circles (labeled "$a_1$", "$b_1$", "$a_2$", and "$b_2$"), obtained via the Maxwell construction. The spinodal points are marked by filled circles (labeled "$c_1$", "$d_1$", "$c_2$", and "$d_2$"). Note that the reentrant curve is symmetric with respect to the horizontal line corresponding to $\mu = -\varepsilon_1 = -1$ (red dashed line). In panel (b): an illustration of the effect of the demixing tendency between two poor solvents on polymer chains ($\chi_s$) on the square of the external adsorption field, $U^2$.

From **Figure 2** to **Figure 4**, we also observe that there exist non-plain spinodal points that depend on the grafting density. To obtain these non-plain spinodal points and simplify our calculations, we consider small values of the monomer concentration by expanding the *c*-dependent terms in the logarithmic function of **Equation(11)** up to quadratic accuracy ($c^2$, i.e., the second-virial expansion). This additional constraint for the phase transition of polymer brushes can typically be satisfied when the grafting density is not too high, as can be seen from the examples shown in **Figure 2b** and **Figure 3b**. This leads to the simplification of **Equation (11)** to

$$U^2 = \left(1 - \frac{1}{2}\chi_s - y\right)^2 \left(k - \frac{s^2}{y^3}\right) \tag{13}$$

Here, we introduce the rescaled order parameter ($c \to y$), the rescaled grafting density ($\sigma \to s$), and the parameter $k$ as

$$\begin{aligned} y &= (\varepsilon_2 + \varepsilon_3) c \\ s &= (\varepsilon_2 + \varepsilon_3) \sigma \\ k &= \frac{(1 + \varepsilon_2 - \varepsilon_3)}{2(\varepsilon_2 + \varepsilon_3)} \end{aligned} \tag{14}$$

Minimizing $U^2$ with respect to $y$, i.e., $\partial U^2/\partial y = 0$ using a leading-order approximation for the variable $y$, yields the rescaled monomer concentration at the onset of spinodal decomposition for a polymer brush

$$y_1 \approx \left[\frac{3}{2k}\left(1 - \frac{1}{2}\chi_s\right) s^2\right]^{\frac{1}{4}} \tag{15}$$

Inserting **Equation(15)** into **Equation(13)** leads to the spinodal relation in the phase diagram of parameter space for low brush density

$$\frac{(\mu + \varepsilon_1)^2}{8} \approx \left(1 - \frac{1}{2}\chi_s\right)^2 k - \left(1 - \frac{1}{2}\chi_s\right)^{\frac{5}{4}} \left(\frac{2}{3} k\right)^{\frac{3}{4}} \sqrt{s} \tag{16}$$

The resulting phase diagram is shown in **Figure 5**. **Equation(16)** can be used to construct corresponding phase diagrams for the cosolvency effect in polymer brushes based on previous experimental data **[17, 33-36]**. **Equation (16)** prescribes that an increasing of demixing tendency ($\chi_s$) between cosolvent and solvent leads to a stronger phase transition, i.e., a more pronounced cosolvency effect for the same polymer, provided other model parameters are fixed (see **Figure 5** for details). We note that this prediction is consistent with previously known theories **[19-23]** on the cosolvency effect. The existence of a physically meaningful solution to **Equation (16)** imposes the following upper bound on the grafting density

$$\sigma_{u1} = \frac{3\sqrt{3}}{4}\sqrt{1+\varepsilon_2-\varepsilon_3}\left(\frac{1-\frac{1}{2}\chi_s}{\varepsilon_2+\varepsilon_3}\right)^{\frac{3}{2}} \tag{17}$$

subject to the physical constraint that $1 + \varepsilon_2 > \varepsilon_3$.

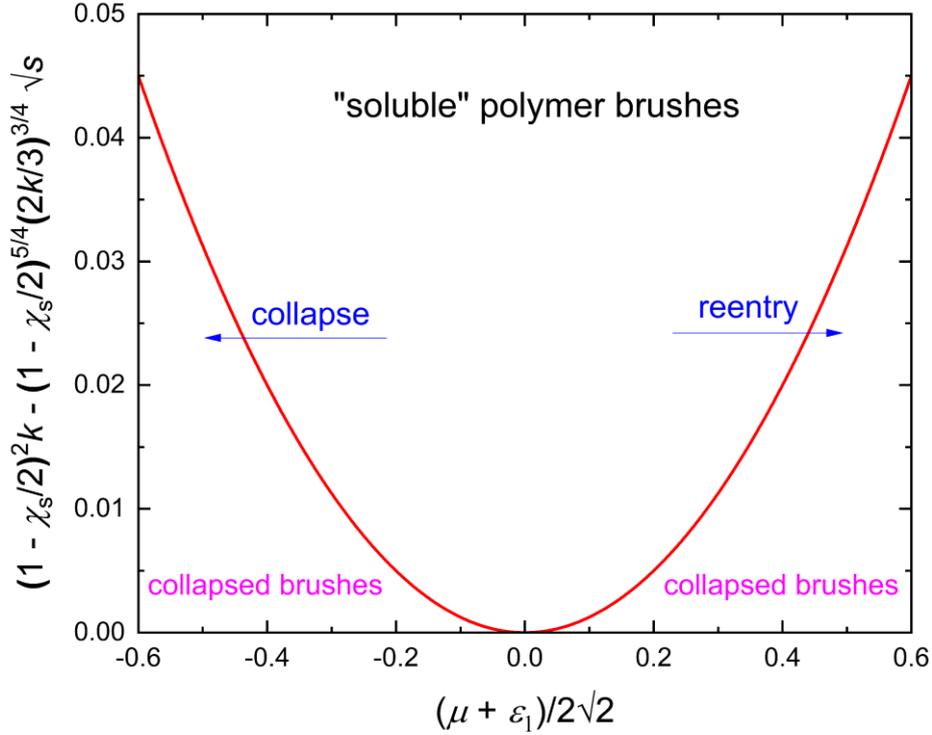

**Figure 5.** Phase diagram of parameter space for the discontinuous phase transition of polymer brushes toward the collapsed state, as described by **Equation (16)**, obtained within the approximation of low brush density.

Equation(16) is based on a second-order approximation for monomer concentration (*c*) in **Equation(13)**. This approximation will break down at higher concentrations, i.e., at higher grafting densities. To estimate the upper bound of the grafting density, we consider an expansion of the *c*-dependent terms in the logarithmic function of **Equation(11)** up to the cubic order ($c^3$). Now, to obtain the spinodal for the instability of the collapsed state toward swelling, we neglect the brush elasticity of the high-density phase. With this approximation, we obtain a truncated function $U^2(c)$ from **Equation(11)** as

$$U^2(c) = \left(1-\frac{1}{2}\chi_s-y\right)^2\left[k+\frac{y}{3(\varepsilon_2+\varepsilon_3)^2}\right] \tag{18}$$

The condition of $\partial U^2/\partial y = 0$ leads toward the position of the local maximum of **Equation(18)** as

$$y_2 \approx \frac{1}{3} - \frac{1}{6}\chi_s - (1 + \varepsilon_2 - \varepsilon_3)(\varepsilon_2 + \varepsilon_3) \tag{19}$$

A rough estimate for the upper grafting density can be obtained by equating the two values for the points of instability, i.e., $y_1 = y_2$. Using the result of **Equation(15)**, another upper grafting density ($\sigma_{u2}$) is given by

$$\sigma_{u2} \approx \frac{\sqrt{3}}{3}\left[\frac{1 - \frac{1}{2}\chi_s}{3(\varepsilon_2 + \varepsilon_3)} - (1 + \varepsilon_2 - \varepsilon_3)\right]^2 \sqrt{\frac{(\varepsilon_2 + \varepsilon_3)(1 + \varepsilon_2 - \varepsilon_3)}{1 - \frac{1}{2}\chi_s}} \tag{20}$$

with the physical constraint of $1 + \varepsilon_2 > \varepsilon_3$. By **Equation (17)** and **Equation (20)**, we have

$$\frac{\sigma_{u2}}{\sigma_{u1}} \approx \left[\frac{2}{9} - \frac{2(\varepsilon_2 + \varepsilon_3)(1 + \varepsilon_2 - \varepsilon_3)}{3\left(1 - \frac{1}{2}\chi_s\right)}\right]^2 \tag{21}$$

By **Equation (21)**, we see that $\sigma_{u2} < \sigma_{u1}$ for the condition of $\varepsilon_2 + \varepsilon_3 \leq 1 - \chi_s/2$ with $0 < \varepsilon_2 < 1/2$ and $\varepsilon_3 > 1/2$. However, we have $\sigma_{u2} > \sigma_{u1}$ for the condition of $\varepsilon_2 + \varepsilon_3 \gg 1 - \chi_s/2$ with $1/2 < \varepsilon_3 < 1$.

Above $\sigma_u$ only a smooth crossover can take place for the variation of the cosolvent density, such as displayed on **Figures 2a** for cases of $\sigma > 0.15$. This condition limits the validity of the phase diagram in **Figure 5**, i.e., only the points ($\sigma$, $\varepsilon_3 - \varepsilon_2$) below the critical condition can be considered for the spinodal line. We note that the full phase diagram is three-dimensional ($\mu + \varepsilon_1$, $\varepsilon_3 - \varepsilon_2$, $\sigma$), and the **Equation (16)** is only valid within the second-virial expansion to allow a simplification to the 2D phase diagram. The critical condition can be used to construct the 3D phase diagram for the spinodal by adding of the third axis given by $\varepsilon_3 - \varepsilon_2$. Then, the spinodal becomes a surface which extends up to a limiting grafting density given by the critical condition for a given value of $\varepsilon_3 - \varepsilon_2 < 1$. We recognize that the analytical approximation in this section is only a sketch of the general properties of the phase diagram. A detailed phase diagram can be constructed in principle via a combination of using the Maxwell construction and numerically calculating the critical point of squared external adsorption field $U^2$ by means of

$$\frac{\partial(U^2)}{\partial c} = \frac{\partial^2(U^2)}{\partial c^2} = 0 \tag{22}$$

However, a detailed investigation of this aspect lies beyond the focus of this work and deserves further study in the future.

## 5. Connections with polymer solutions

The general physics underlying the cosolvent-induced discontinuous phase transition in the cosolvency effect can be understood by considering the thermodynamic equilibrium between polymer brushes and the cosolvent when the brush chains are infinitely long ($N \to \infty$) and the grafting density is infinitely low (but nonzero). We recognize that this situation is thermodynamically equivalent to the absence of translational entropy of long polymer chains, i.e., neglecting the term $(c/N)\ln(c)$ in the free energy of the Flory–Huggins solution theory **[24, 25]**. The free energy per monomer unit of the polymer solution, $g_s$, can be denoted according to **Equation (8)** by ignoring constant terms as

$$\begin{aligned} g_s &= \frac{1}{2}(\varepsilon_2 - \varepsilon_3)c - \frac{1}{8}\frac{(\mu + \varepsilon_1)^2}{\left[1 - \frac{1}{2}\chi_s - (\varepsilon_2 + \varepsilon_3)c\right]} + \left(\frac{1}{c} - 1\right)\ln(1-c) \\ &= v_1 + \frac{1}{2}v_2 c + \frac{1}{6}v_3 c^2 + \frac{1}{12}v_4 c^3 + O(c^4) \end{aligned} \tag{23}$$

with the virial-expansion coefficients up to the fourth order of monomer concentration

$$\begin{aligned} v_1 &= -\frac{1}{8}\frac{(\mu + \varepsilon_1)^2}{\left(1 - \frac{1}{2}\chi_s\right)} - 1 \\ v_2 &= 1 + \varepsilon_2 - \varepsilon_3 - \frac{(\varepsilon_2 + \varepsilon_3)}{4\left(1 - \frac{1}{2}\chi_s\right)^2}(\mu + \varepsilon_1)^2 \\ v_3 &= 1 - \frac{3(\varepsilon_2 + \varepsilon_3)^2}{4\left(1 - \frac{1}{2}\chi_s\right)^3}(\mu + \varepsilon_1)^2 \\ v_4 &= 1 - \frac{3(\varepsilon_2 + \varepsilon_3)^3}{2\left(1 - \frac{1}{2}\chi_s\right)^4}(\mu + \varepsilon_1)^2 \end{aligned} \tag{24}$$

Here, the $i$-th viral coefficient is denoted as $v_i$.

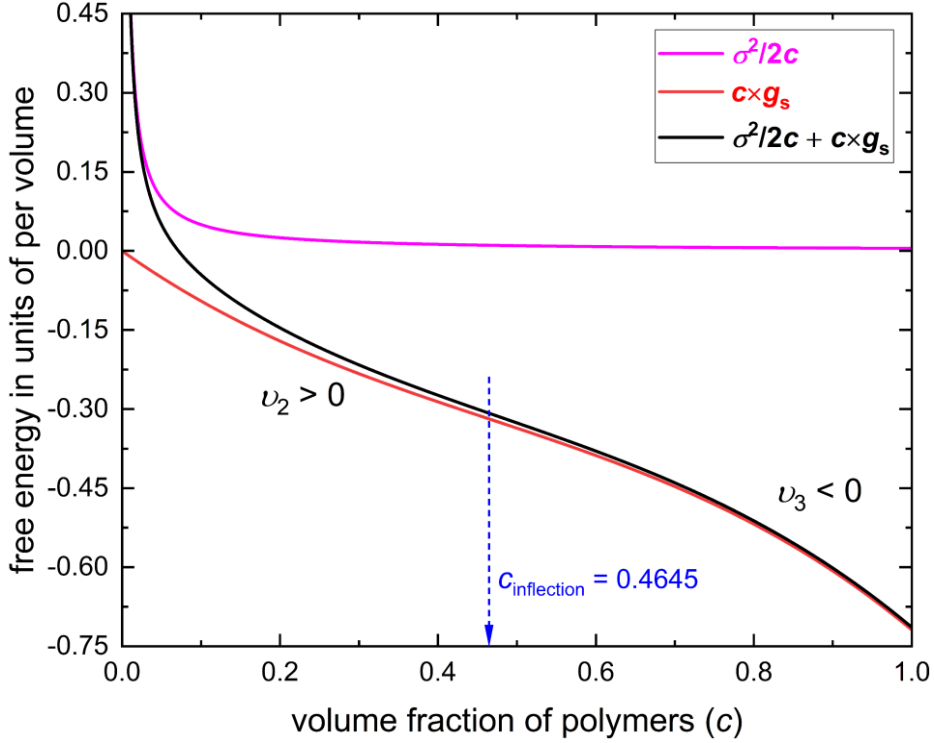

**Figure 6.** An example of the free energy per unit volume as a function of polymer concentration for the situation where the third virial coefficient is negative, but the second virial coefficient is positive (see **Equation (24)**). The model parameters used in the figure are $\sigma = 0.1$, $\varepsilon_1 = 1.0$, $\varepsilon_2 = 2.5$, $\varepsilon_3 = 0.55$, $\chi_s = 0.5$, and $U = 0.6147$ (corresponding to $\mu = -0.3853$). The second ($v_2$) and third ($v_3$) virial coefficients are respectively given by 2.4379 and $-5.2479$. The inflection point of the free energy is $c_{\text{inflection}} = -v_2/v_3 = 0.4645$. The elastic free energy of the polymer brush (pink line) shifts the discontinuity, or depending on the grafting density completely suppresses the discontinuity.

The critical point of the free energy per unit volume ($cg_s$) is determined by $d^2(cg_s)/dc^2 = d^3(cg_s)/dc^3 = 0$. This is roughly equivalent to keep the second virial coefficient ($v_2$) positive but the third viral coefficient ($v_3$) stays negative, a phase-transition feature which was first denoted by de Gennes **[55]** as type II phase transition to particularly explain the lower critical solution behavior **[56]** of uncharged polymers. This situation is given by

$$\frac{\left(1-\frac{1}{2}\chi_s\right)^3}{3\left(\varepsilon_2+\varepsilon_3\right)^2} < \left(\frac{\mu+\varepsilon_1}{2}\right)^2 < \left(\frac{1+\varepsilon_2-\varepsilon_3}{\varepsilon_2+\varepsilon_3}\right)\left(1-\frac{1}{2}\chi_s\right)^2 \qquad (25)$$

An example of **Equation (25)** is displayed on **Figure 6**. The elastic contribution of polymer brushes to the free energy per unit volume, $\sigma^2/2c$, suppresses states with low concentrations, as illustrated in **Figure 6**. This shifts the transition toward lower values of the control parameter $|\mu + \varepsilon_1|$, but can also completely suppress the discontinuity for high grafting densities, resulting in a smooth crossover. The highest possible external field is given by $(\mu + \varepsilon_1)^2 \approx 4(1+ \varepsilon_2 - \varepsilon_3)(1 - \chi_s/2)^2/(\varepsilon_2 + \varepsilon_3)$, see **Equation (25)**, which is a moderate condition that does not violate the perturbation expansion introduced by **Equation(5)**.

Using **Equation (25)**, we can determine a necessary window for the parameter $\varepsilon_2$ within which a discontinuous phase transition can occur, provided that $\varepsilon_3 > 1/2$ is already known. This leads to

$$\varepsilon_2 > -\frac{1}{2} + \sqrt{\left(\varepsilon_3 - \frac{1}{2}\right)^2 + \frac{1}{3}\left(1 - \frac{1}{2}\chi_s\right)} \qquad (26)$$

Thus, the discontinuous collapse/reentry transition can be traced back to a sign change in the effective three-monomer interaction, $v_3$, while $v_2$ remains positive. This is induced by the repulsion coupling ($\varepsilon_2$) between monomers due to the adsorbed cosolvent, which prescribes a necessary strength for $\varepsilon_2 > 0$ to initiate cosolvency, as given by **Equation (26)**. It is important to note that it is not the effective two-monomer interaction that causes the discontinuity, as one might expect from the classical picture of "dislike" between polymer chains. A change in the sign of $v_2$ alone merely results in a continuous, $\theta$-point-like transition for polymer brushes **[57, 58]**. This emphasizes the role of the nonlinear $\chi$-function and justifies the notion of a "second type" (type II) of transition. We recognize that the condition in **Equation (26)** becomes trivial if the right-hand side is non-positive for some parameter values of $\varepsilon_3 > 1/2$ and $\chi_s > 0$. This implies that additional conditions must be identified to determine the occurrence of the phase transition (cosolvency). We will uncover these conditions in the next section.

## 6. The role of cosolvent-solvent interaction in the bulk

The analytic results in previous sections predicted the existence of a cosolvency phase transition for polymer brushes with no restriction on the minimum value of the preferential-solvation strength ($\varepsilon_1$). This is clearly not the case, as it is expected that there is no cosolvency effect when there is no preferential solvation ($\varepsilon_1 = 0$). In other words, the threshold value of the preferential-solvation strength ($\varepsilon_1$) should play a role in the occurrence of the cosolvency effect; this was clearly implied by both previous experimental **[9-11, 14, 17, 18, 28, 30-32]** and simulation **[26-29]** studies. In this section, we determine the minimum preferential-solvation strength ($\varepsilon_{1,\,min}$) required for a phase transition to occur.

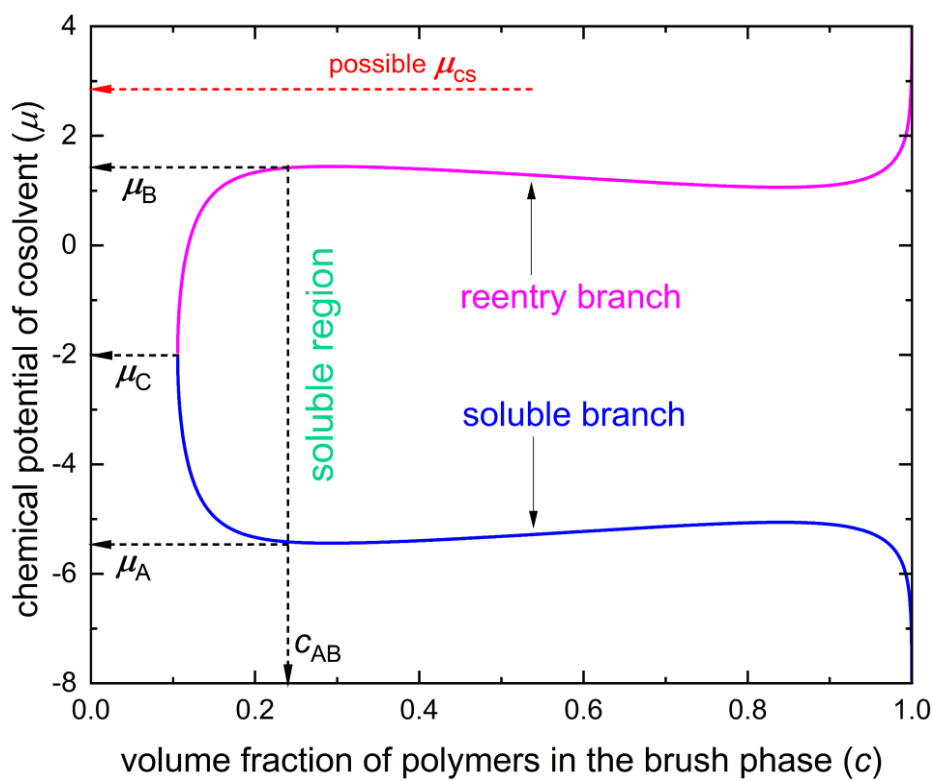

**Figure 7.** Definition of the soluble transition and reentry transition. In the figure, $\mu_A$ denotes the cosolvent chemical potential at the onset of the soluble transition, $\mu_B$ denotes the cosolvent chemical potential at the onset of the reentry transition, $\mu_C$ denotes the chemical potential at which the polymer brush exhibits maximum swelling, and $\mu_{cs}$ denotes the chemical potential at which the cosolvent begins to preferentially form cosolvent-solvent clusters.

Besides the preferential adsorption of cosolvent to polymer chains, a cosolvent molecule may also form cosolvent-solvent clusters in the bulk. This is well known in alcohol-water mixtures, where an alcohol-water cluster is a relatively stable supramolecular structure **[59]**. Thus, the formation of cosolvent-solvent clusters can further compete with and regulate whether polymers exhibit the cosolvency effect. The exchange chemical potential ($\mu_{cs}$) for forming a cosolvent-solvent cluster in our study can be roughly estimated in a similar way to our previous works **[45, 46]** by

$$\mu_{cs} \approx -\frac{1}{2}\varepsilon_{cs} \tag{27}$$

Here, the numerical prefactor 1/2 is not an exact but a rough estimate. In **Equation (27)**, $\varepsilon_{cs}$ is the average interaction strength of the cosolvent-solvent interaction in the formation of a cosolvent-solvent cluster. The sign of $\varepsilon_{cs}$ is positive for attractive interaction and negative for repulsive interaction. For example, in alcohol-water mixtures, $\varepsilon_{cs}$ is positive and its magnitude is on the order of the strength of a hydrogen bond **[60, 61]**. We define several cosolvent chemical potentials to facilitate our following discussions. As shown in **Figure 7**, $\mu_A$ is defined as the cosolvent chemical potential at the onset of the soluble transition, $\mu_B$ as that at the onset of the reentry transition, and $\mu_C$ as the chemical potential at which the polymer brush exhibits the maximum extent of swelling.

Because our model is symmetry, we have the following exact and simple relations

$$\begin{aligned}\mu_C &= -\varepsilon_1 \\ \mu_C &= \frac{1}{2}(\mu_A + \mu_B)\end{aligned} \tag{28}$$

According to **Equation (16)**, we also have the following approximations for low grafting-density brushes

$$\begin{aligned}\mu_A &\approx -2\sqrt{2}\left(1-\frac{1}{2}\chi_s\right)^{\frac{5}{8}}\left[\left(1-\frac{1}{2}\chi_s\right)^{\frac{3}{4}}k-\left(\frac{2}{3}k\right)^{\frac{3}{4}}\sqrt{s}\right]^{\frac{1}{2}} -\varepsilon_1 \\ \mu_B &\approx 2\sqrt{2}\left(1-\frac{1}{2}\chi_s\right)^{\frac{5}{8}}\left[\left(1-\frac{1}{2}\chi_s\right)^{\frac{3}{4}}k-\left(\frac{2}{3}k\right)^{\frac{3}{4}}\sqrt{s}\right]^{\frac{1}{2}} -\varepsilon_1\end{aligned} \tag{29}$$

A basic prediction of **Equation (29)** is that the concentration range of cosolvent for the soluble state of polymer brushes (the soluble region in **Figure 7**, i.e., $|\mu_B - \mu_A|$), becomes narrow when the demixing tendency ($\chi_s$) between the two poor solvent increases provided that other model parameters are fixed. It is remarkable that this prediction concurs with our previous experimental observation **[11]** on cosolvency of a copolymer in a series of binary mixtures of alcohols and water.

Obviously, as illustrated in **Figure 7** when $\mu_{cs} > \mu_B$, there is a cosolvency effect (a full reentrant transition including of both soluble and reentry transitions) due to the preferential solvation of polymer chains by cosolvent. This situation corresponds to

$$\varepsilon_1 > 2\sqrt{2}\left(1-\frac{1}{2}\chi_s\right)^{\frac{5}{8}}\left[\left(1-\frac{1}{2}\chi_s\right)^{\frac{3}{4}}k-\left(\frac{2}{3}k\right)^{\frac{3}{4}}\sqrt{s}\right]^{\frac{1}{2}}+\frac{1}{2}\varepsilon_{cs} \qquad (30)$$

Provided that $\varepsilon_3 > 1/2$ is already known, the insertion of **Equation (26)** into **Equation (30)** leads to a sufficient estimation

$$\varepsilon_1 > \frac{1}{2}\varepsilon_{cs}+(2-\chi_s)\sqrt{\frac{1+\varepsilon_2-\varepsilon_3}{\varepsilon_2+\varepsilon_3}} > \frac{1}{2}\varepsilon_{cs}+\sqrt{6(2-\chi_s)}\left[\sqrt{\left(\varepsilon_3-\frac{1}{2}\right)^2+\frac{1}{3}\left(1-\frac{1}{2}\chi_s\right)}-\left(\varepsilon_3-\frac{1}{2}\right)\right] \qquad (31)$$

The right-hand side of **Equation (31)** is a monotonic decreasing function with respect to parameters $\varepsilon_3 > 1/2$ and $\chi_s < 2$. **Equation (31)** prescribes that the minimum strength of preferential solvation ($\varepsilon_{1,\,min}$) to initiate cosolvency effect is about $\varepsilon_{1,\,min} \approx 2k_BT$, provided that the strength of cosolvent-solvent interaction $\varepsilon_{cs}$ is attractive **[60, 61]** and on the order about 1 $k_BT$ with $1/2 < \varepsilon_3 < 1$. We note that the preferential-solvation strength of $\varepsilon_{1,\,min} \approx 2k_BT$ is a normal prescription for the strength of anisotropic interactions **[62]** such as hydrogen bonds and dipole interactions.

We can also determine a necessary condition for the cosolvency effect to occur for a polymer as

$$\varepsilon_1 > \frac{1}{2}\varepsilon_{cs} \qquad (32)$$

This result is obtained by setting the first term of **Equation (30)** to zero, which is possible by choosing suitable values for the parameters $\varepsilon_2$, $\varepsilon_3$, $\chi_s$, and $\sigma$. Note that here the cosolvent-solvent interaction ($\varepsilon_{cs}$) in bulk can be ether attractive ($\varepsilon_{cs}$ with positive values) or repulsive ($\varepsilon_{cs}$ with negative values). We note that this analytical prediction is consistent with the cosolvency effect observed for polymers in alcohol-water mixtures [9, 11], where the interaction between alcohol and water can be either attractive or repulsive [60, 61]. We recognize that the strength of $\varepsilon_{cs}$ is typically on the order of 1 $k_B T$ for alcohol-water mixtures [60, 61], which implies $\mu_C = -\varepsilon_1 \approx \varepsilon_{cs}/2$. If we simply assume that the exchange chemical potential $\mu$ depends only on the volume fraction $x$ of the cosolvent in the bulk, for example, $\mu \approx \ln(x) - \ln(1-x)$, this suggests that the volume fraction of cosolvent at maximum swelling of polymer brushes is approximately $x_c \approx 0.35$. Remarkably, this analytical prediction is not far from experimental reports on the cosolvency effect in polymer brushes [35, 36]. According to **Equation (31)** and **Equation (32)**, we see that a repulsive cosolvent-solvent interaction in the bulk is not necessary for the cosolvency effect to occur. We note that our result differs qualitatively from the pioneering theoretical formulations reported by Meng [19, 21], Zhang [20], Xiao [22], and Dudowicz [23] with their respective coworkers, who showed that a repulsive interaction between two poor solvents in the bulk is necessary for the cosolvency effect to occur for a polymer.

## 7. Discussion

One motivation for this work stemmed from the observation that while experimental studies [17, 33-36] have reported on the cosolvency effect in polymer brushes, no analytic theory has yet been developed to describe it. A second motivation arose from the fact that both molecular dynamics simulations [26-29] and experiments [9-11, 14, 17, 18, 28, 30-32] have indicated that the cosolvency effect can arise due to the relatively preferential solvation of polymer chains by the cosolvent. However, no analytic theory has considered the role of preferential solvation in determining the phase behavior of cosolvency. Therefore, the primary goal of this work is to formulate the first analytic theory that explores the cosolvency effect in polymer brushes. We systematically analyze the role of preferential solvation in cosolvency from a theoretical perspective. Based on this analysis, we propose a mean-field model that incorporates only the essential physics yet captures the experimentally observed trends.

We show that the relatively preferential solvation by cosolvent yields an effective repulsion coupling between cosolvent-solvated and solvent-solvated monomers of polymer brushes. The equilibrium solvation of polymer chains by cosolvent leads to a concentration-dependent $\chi$-function, which describes effective interactions within the brush and reproduces the reentrant feature of the cosolvency effect. This framework predicts a discontinuous soluble transition followed by reentrant re-collapse at higher cosolvent concentrations. For low-density brushes, we derive an analytical approximation and identify the phase diagram of parameter space for discontinuous transitions. To simplify the analytical solution of our theory, we considered in detail the case of two symmetric poor solvents, i.e., $\varepsilon_3 = \varepsilon_4 > 1/2$. This analytical treatment within a minimal free-energy model reveals that swelling and re-collapse transitions share the same thermodynamic signature, a result that however need not hold for the asymmetric case $\varepsilon_3 \neq \varepsilon_4$ with $\varepsilon_3 > 1/2$ and $\varepsilon_4 > 1/2$. Asymmetric behavior in the cosolvency response of polymer brushes can also be analyzed analytically. This can be addressed by introducing a perturbation $\delta$ defined by $\varphi = t(1 - \delta)$, where $t \equiv (\varepsilon_2 + \varepsilon_4)/(2\varepsilon_2 + \varepsilon_3 + \varepsilon_4)$, as can be seen from the constructions of $g_{rep}$ and $g_{attr}$ in our free-energy model. It is readily observed that maximum coupling is achieved at $\varphi = t = (\varepsilon_2 + \varepsilon_4)/(2\varepsilon_2 + \varepsilon_3 + \varepsilon_4) < 1$, where the maximum expansion of polymer chains is expected. Following the same step-by-step calculation procedure as in previous sections, one can obtain corresponding analytic results. However, these expressions for the asymmetric case are cumbersome and do not alter the essential physical conclusions of the theory; they are therefore not presented in this work for simplicity.

Our theoretical approximations base on the simplest free energy model of polymer brushes, which assumes a homogeneous density profile in accordance with the classical Alexander–de Gennes approach [39, 40]. Using this simple approach, our theory predicts that a discontinuous soluble transition in the homogeneous model can lead to vertical phase separation, resulting in a type II coexistence state within the same brush. However, due to the current lack of experimental and simulation data on polymer brushes [17, 33-36], this prediction awaits future confirmation. We note that the cosolvency effect is the antisymmetric counterpart of the cononsolvency effect, a well-known reentrant condensation phenomenon in which a polymer is poorly soluble in mixtures of two good solvents. By analogy with the seminal works on cononsolvency by Sommer and coworkers [48, 63-67], who also employed the classical Alexander–de Gennes approach [39, 40], we anticipate that future experimental or simulation evidence supporting the discontinuous soluble transition predicted here for cosolvency could also validate the mapping between cosolvency and the type II transition. A detailed test of our model could be carried out through computer simulations and experimental studies by varying parameters such as grafting density, brush chain length, and the types of cosolvent and solvent.

Our theory for cosolvency of polymer brushes can be refined and extended in several directions. One interesting way would be to modify our model to incorporate a nonhomogeneous density profile, such as the so-called parabolic profile **[68, 69]** and the polydisperse effect **[70, 71]**, which may have nontrivial consequences such as a possible sharper phase transition regarding cosolvent concentration **[65]**. Under the context of the Alexander–de Gennes approach **[39, 40]**, we note that this issue can be understood by refining our theory with a "polydisperse model" **[72]** for chain elasticity that proposed by Smook, de Beer, and Angioletti-Uberti for an arbitrary brush chain length distribution. Another intriguing extension would involve examining how cosolvency behavior is affected by the addition of small components with different underlying mechanisms. For instance, the depleted depletion mechanism **[73]** reported in all-atom computer simulations, which could interfere with cosolvent adsorption (regulatory solvents or solutes) or influence the thermal mixing behavior of polymers. However, a detailed investigation of these aspects lies beyond the scope of this work and deserves further study in the future.

## 8. Summary

To conclude, in constructing the theory, we focused on the solvation, repulsion, and attraction effects of the two solvents near or on polymer chains arising from anisotropic interactions, while neglecting their non-essential mixing effects when solvent molecules are far from the polymer chains. Using this approach, we have demonstrated that the essential features of the cosolvency response in polymer brushes can be rationalized within the framework of relatively preferential solvation by the cosolvent, which gives rise to an effective repulsive coupling between cosolvent-solvated and solvent-solvated monomers. For cosolvency to occur, our theory prescribes a minimum strength for the effects of preferential solvation and the associated repulsive coupling. Our theory shows that an increase in the demixing tendency between cosolvent and solvent can lead to a stronger phase transition (i.e., a more pronounced cosolvency effect for the same polymer), in agreement with previous theories **[19-23]**. However, our theory also demonstrates that a repulsive cosolvent–solvent interaction in the bulk is not necessary for the cosolvency effect to occur, a conclusion that qualitatively differs from earlier theoretical work **[19-23]**.

Meanwhile, our theory delineates a phase diagram of parameter space within which a first-order phase transition can emerge due to the cosolvency effect in polymer brushes, as formalized in **Equation (16)** and illustrated in **Figure 5** which has not been previously reported. From the perspective of practical applications, this finding suggests that stimulus-responsive behavior in polymer brushes can be achieved by harnessing the sudden change in brush thickness associated with cosolvency. Such tunable swelling and collapse transitions offer promising avenues for designing adaptive surface coatings **[34, 35]**. By providing a predictive analytical framework, this work lays the groundwork for the rational design of smart stimulus-responsive materials based on the cosolvency effect in polymer brushes, a capability which was not established to date.

## Author Information

**Corresponding Author**

Huaisong Yong, yonghs@swpu.edu.cn, yonghuaisong@gmail.com

**Notes**

The authors declare no competing financial interest.

## Acknowledgments

Huaisong Yong acknowledges the financial support for this research from the "Tianfu Emei" Scholar Foundation of Sichuan Province (No. 2326) and the visiting project licensed by the Federal Office for Migration and Refugees of Germany (No. BAMF 5-00002915). Huaisong Yong also acknowledges that the numerical calculations in this research were performed on the supercomputing system at the High-Performance Computing Center of Southwest Petroleum University. The authors thank Holger Merlitz and Jens-Uwe Sommer at the Leibniz-Institut für Polymerforschung Dresden, Sissi de Beer at the University of Twente, and Chuanfu Luo at Changchun Institute of Applied Chemistry (Chinese Academy of Sciences) for the fruitful exchanges of opinions related to this research.